\documentclass[aps,pra,twocolumn,amsmath,
showpacs]{revtex4}
\usepackage{graphicx}
\begin{document}
\title{Cluster states of Fermions in the single $l$-shell model}
\author{Andr\'as Csord\'as}
\email{csordas@tristan.elte.hu}
\affiliation{Research Group for Statistical Physics of the 
             Hungarian Academy of Sciences,
             P\'azm\'any P. S\'et\'any 1/A, H-1117 Budapest, Hungary}
\author{\'Eva Sz\H oke}
\email{jeva@ludens.elte.hu}
\affiliation{Department of Physics of Complex Systems, E\"otv\"os University,
             P\'azm\'any P. S\'et\'any 1/A, H-1117 Budapest, Hungary}
\altaffiliation{The work has been done while beeing a member of the
             teaching staff of the
             Budapest University Catholic Secondary School,
             Szab\'o Ilonka u. 2-4, H-1015 Budapest, Hungary}
\author{P\'eter Sz\'epfalusy}
\email{psz@complex.elte.hu}
\affiliation{Department of Physics of Complex Systems, E\"otv\"os University,
             P\'azm\'any P. S\'et\'any 1/A, H-1117 Budapest, Hungary}
\affiliation{Research Institute for Solid State Physics and Optics,
             P. O. Box 49, H-1525 Budapest, Hungary}
\date{\today}
\begin{abstract}
The paper concerns the ground state structure of the partly filled $l$-shell
of a fermionic gas of atoms of spin $s$ in a spherically 
symmetric spin independent trap potential. At particle
numbers $N=n(2s+1)$, $n=1,2,\ldots,2l+1$ the basic building blocks
are clusters consisting of $(2s+1)$ atoms, whose wave functions
are completely symmetric and antisymmetric in space and spin variables,
respectively. 
The creation operator of a cluster is constructed whose repeated
application to the vacuum leads to the multi-cluster state.
Ground state energy expressions are derived for the
$n$-cluster states at different $l,s$ values and interpreted in simple terms.
\end{abstract}
\pacs{05.30.Fk,31.15.Hz,21.60.Cs,74.20.Fg}
\maketitle

\section{Introduction}

The many body problem as applied to finite systems has a long history
in atomic and nuclear physics \cite{Fano59,Judd67,Szasz92,Ring04}. 
One of the central problems has been the
nature of the ground state in case of a partially filled shell.

When we consider atoms (a gas at zero temperature)
in an external potential the possible
behaviors are quite rich. One can assume that the collision between
atoms does not excite internal degrees of freedom and the atoms can be 
regarded as structureless objects whose spins are fixed being in a 
definite hyperfine state. The external potential can be supplied by a
magnetic or optical trap \cite{Pethick04}. 
In the past few years a very intensive research
has been continued in case of Bose-particles both theoretically and
experimentally in a variety of such systems. Fermion systems along these
lines have been less studied until recently, but important achievements
have been already available and one can be sure that rapid development will
continue in the future. In particular the achievement and study of the
superfluid state has become one of the frontiers in physics
(see for Reviews \cite{Stoof99,Pethick04a,Chen05,Heiselberg04}, which 
contain references to earlier works).

In this paper we treat fermionic systems. The external potential will be
assumed to be spherically symmetric and the interaction between atoms
will be described by a spin independent $\delta$ function like 
attractive pseudopotential. Within the single $l$-shell model of fermionic
atoms with arbitrary large spins we study the ground state properties of
the gas.

We concentrate on cluster states containing particles of numbers 
$N_n=n(2s+1)$ where $s$ is the spin of the particles and 
$n=1,2,\ldots,2l+1$. The existence of such clusters have been pointed
out by us previously \cite{Csordas04}. In the present paper we extend
the investigations in several directions. It is advantageous to use
the second quantized representation, which makes possible, among other
things, to obtain the multi-cluster states by repeated applications of
of the cluster creation operator to the vacuum.
Besides numerical calculations extensive analytical
studies are carried out. It is shown that 
to a good approximation the
hamilton operator can be replaced by a model one built of the the
operator of the particle number, of the quadratic Casimir operators of the
groups $SU(2l+1)$ and $SO(2l+1)$. It is shown that within the model
the clusters can be 
conceived as interacting (Cooper) pairs if $s>1/2$, the interaction
being of statistical origin. The model Hamiltonian coincides
with the true one for $l=1,2$ (The special cases mainly investigated
in \cite{Csordas04}).

For the model hamiltonian the cluster states are explicitly given and
it is pointed out that the energy of the $n$-cluster state can be
written as the sum of energies of the clusters proportional to $n$
and an ``interaction term'' term between the clusters proportional to
$n(n-1)/2$. It is found that the second term disappears when $s=1/2$.
For $s=1/2$ it was shown by Racah in 1952 \cite{Racah52} that
the ground state consists of independent pairs.

Investigations of particles in an open shell has been
an important area in atom and nuclear physics. 
As discussed above in the trapped gas
of Fermi atoms new features appear due to the fact that the spin of
the particles can be higher than $1/2$. Comparing with the situation
in the atom shell a further important difference is that instead of the
long ranged Coulomb force between the electrons the atom-atom interaction
which has to be considered is short ranged, while comparing the situation
with that of an open neutron shell in a nucleus an important difference is 
that in our case the open shell is an $l$-one (since no-spin orbit 
interaction is present) as contrasted to the $j$-one in the nucleus.

Though our main purpose in this paper is to enlarge the picture we have
about the dynamics of fermionic particles in a partly filled shell, a 
few words about the relevance of the model for physically realizable 
situations in case of trapped gases are in order. 
Obviously optical traps are the suitable ones which allow the free rotation
of the spins. It is assumed that the external potential contains besides
the confining potential (which is typically of a harmonic oscillator type)
the mean field of the atoms building the closed shells. There are two
conditions then to be fulfilled in order that the single-$l$ shell model apply.
Firstly, the mean field due to the closed shells should be strong enough
that the possible degeneracy of the levels of different $l$ values be
lifted considerable. Secondly, the characteristic interaction energy 
of the atoms in the partly occupied shell should be smaller than the
relevant level distance in the external potential. Both requirements
can be satisfied simultaneously if the strength of the interaction
between the atoms is weak and the number of the atoms in the trapped
gas is sufficiently large \cite{bruun02a,bruun02}.

The paper is organized as follows.
In Section \ref{sec:Formulation} the Hamiltonian is written in terms
of irreducible tensor operators and a model Hamiltonian is introduced
built from the operator of particle number and from the quadratic
Casimir operators of the groups $SU(2l+1)$ and $SO(2l+1)$. The Section
\ref{sec:clust_states} is devoted to the extensive investigation of
the one-cluster state. The symmetrizer playing an important role in
creating the cluster states is analyzed in detail in Section 
\ref{sec:symmetrizer}. Energy eigenvalues of the model Hamiltonian
for cluster states generated by the cluster creation operator are
calculated in Section \ref{sec:h_mvec}. It is shown in Section
\ref{sec:aver_ener} that the expectation value of the
Hamiltonian agrees with that of the model one for cluster states.
Section \ref{sec:summary} contains the summary and a discussion of 
the results. Appendices \ref{app:ops} and \ref{app:decomp} present
certain steps of proofs outlined in the main text. Appendix
\ref{app:spinonehalf} discusses the special case when the spin of the 
particles is $1/2$.

\section{Formulation}
\label{sec:Formulation}

Let us assume that the trap potential is spherically symmetric. 
At zero temperature the open shell consists of fermions of spin $s$
which partially fill the $(n,l)$ shell. The one particle normalized
wave functions are given by
\begin{equation}
\Psi_{n,l,m,s,\nu}(r,\vartheta,\varphi,\sigma)=
R_{n,l}(r)Y^l_{m}(\vartheta,\varphi)\chi^s_{\nu}(\varsigma),
\label{eq:onepf}
\end{equation}
Here $\varsigma$ is the discrete spin variable and 
$\chi^s_{\nu}(\varsigma)$ is the normalized spin eigenfunction. 
In the model the quantum numbers $(n,l,s)$ are fixed, $m$ and $\nu$ can take
the values $m=-l,-l+1,\ldots,l$ and $\nu=-s,-s+1,\ldots,s$ and the particle
number $N$ can vary between $0$ and $(2l+1)(2s+1)$.
The functions (\ref{eq:onepf}) are eigenfunctions of the one-particle
Hamiltonian, which contains the kinetic energy term, the trap potential and the
averaged field potential of the closed shells. 
The interaction between the particles in the partially filed shell
is given by
\begin{equation}
H_{int}=-{\lambda \over 2}\sum_{i,j=1 \atop i \ne j}^N
    \delta(\mathbf{r}_i-\mathbf{r}_j),\quad \lambda>0
\label{eq:firstqop}
\end{equation}
corresponding to a spin independent $s$-wave scattering with negative
scattering length. Our aim is to diagonalize (\ref{eq:firstqop})
on the fixed basis (\ref{eq:onepf}).

Let us denote the operator which annihilates a particle with
quantum numbers $(n,l,m,s,\nu)$ by $a_{m,\nu}$. The Hamiltonian 
(\ref{eq:firstqop}) in second quantization reads as
\begin{eqnarray}
\hat{H}_{int}& \equiv  &\hat{h}E_0/\pi  
=-{E_0 \over 2}\sum_{m_1,m_2 \atop m_3,m_4}\sum_{\nu_1,\nu_2}
f_{m_1,m_2;m_3,m_4}
\nonumber \\
&&\times
 a^+_{m_1,\nu_1}a^+_{m_2,\nu_2}a_{m_4,\nu_2}a_{m_3,\nu_1},
\label{eq:genham}
\end{eqnarray}
where $E_0$ is the characteristic energy
\begin{equation}
E_0=\lambda \int_0^\infty |R_{n,l}(r)|^4 r^2\, dr,
\end{equation}
and $f$ can be expressed in terms of the Wigner-$3j$ symbols \cite{Edmonds}
in two equivalent forms
\begin{gather}
f_{m_1,m_2;m_3,m_4}=\int d\Omega\, 
Y^{l*}_{m_1}(\Omega)Y^{l*}_{m_2}(\Omega)Y^{l}_{m_3}(\Omega)Y^{l}_{m_4}(\Omega)
\nonumber\\
=\frac{[l]^2}{4\pi}\sum_{L=0}^{2l}[L]
\left(\begin{array}{ccc}
L & l & l \\
0 & 0 & 0
\end{array}\right)^2 
\nonumber\\
\times\sum_{M=-L}^{L}
\left(\begin{array}{ccc}
l & l & L \\
m_1 & m_2 & -M
\end{array}\right)
\left(\begin{array}{ccc}
l & l & L \\
m_3 & m_4 & -M
\end{array}\right)\label{eq:fmekfirst}\\
=\frac{[l]^2}{4\pi}\sum_{L=0}^{2l}[L]
\left(\begin{array}{ccc}
L & l & l \\
0 & 0 & 0
\end{array}\right)^2 (-1)^{(m_2+m_3)}
\nonumber\\
\times\sum_{M=-L}^{L}
\left(\begin{array}{ccc}
l & l & L \\
m_1 & -m_3 & -M
\end{array}\right)
\left(\begin{array}{ccc}
l & l & L \\
m_4 & -m_2 & -M
\end{array}\right)\label{eq:fmeksecond}
\end{gather}
For notational simplicity we  introduced the symbol $[\ldots]$ 
defined by
\[
[p]\equiv (2p+1).
\]
The dimensionless Hamiltonian $\hat{h}$ (see Eq. (\ref{eq:genham}))
can be written as 
\begin{equation}
\hat{h}=\frac{[l]}{8}\hat{N}-\frac{[l]^2}{8}\sum_{L=0}^{2l}
\left(\begin{array}{ccc}
l&l&L\\
0&0&0
\end{array}\right)^2
\hat{B}_L^2,
\label{eq:hath}
\end{equation}
where 
\begin{equation}
\hat{B}_L^2=\sum_{M=-L}^L (-1)^{L-M} \hat{B}_{L,M} \hat{B}_{L,-M}
\label{eq:blsqr}
\end{equation}
(in eq. (\ref{eq:hath}) only the terms $L$ even remain).
The operators $\hat{B}_{L,M}$ defined as
\begin{eqnarray}
\hat{B}_{L,M}&=&\sum_{m=-l}^l\sum_{\nu=-s}^s (-1)^{l-m}\sqrt{[L]}
\nonumber\\
&& \times
\left(\begin{array}{ccc}
l&l&L\\
m&M-m&-M
\end{array}\right)
a_{m,\nu}^+ a_{m-M,\nu}
\end{eqnarray}
are spin scalars and irreducible tensoroperators with respect
to angular momentum \cite{Fano59}. 
Special cases are $L=0$
\begin{equation}
\hat{B}_{0,0}=\frac{\hat{N}}{\sqrt{[l]}}=\sum_{m,\nu}a_{m,\nu}^+ a_{m,\nu},
\end{equation}
and $L=1$:
\begin{eqnarray}
\hat{B}_{1,0}&=&\frac{\sqrt{3}}{\sqrt{l(l+1)[l]}}\hat{L}_z
\nonumber\\
\hat{B}_{1,\pm 1}&=&\mp\frac{\sqrt{3}}{\sqrt{2l(l+1)[l]}}\hat{L}_{\pm}
\end{eqnarray}
The operator $\hat{B}_{0,0}$ commutes with all the others and
the operators $\hat{B}_{L,M}$ for $L\ge 1$ form  a Lie-group, which is isomorph
to $SU(2l+1)$ with the commutators
\begin{eqnarray}
&&\left[\hat{B}_{L,M},\hat{B}_{L',M'} \right]=
-\sum_{L'',M''}\sqrt{[L][L'][L'']} 
\nonumber\\
&&\times\left[ 1-(-1)^{L+L'+L''} \right] 
(-1)^{M''}
\left(\begin{array}{ccc}
L&L'&L''\\
M&M'&-M''
\end{array}\right)
\nonumber\\
&&\times\left\{\begin{array}{ccc}
L&L'&L''\\
l&l&l
\end{array}\right\}
\hat{B}_{L'',M''}
\end{eqnarray}
($\{\ldots\}$ denotes the Wigner-$6j$ symbol).
Due to the special form of the structure coefficients the operators
$\hat{B}_{L,M}$ for odd $L$ form a subgroup, which is isomorph to
$SO(2l+1)$. This latter also has a subgroup $SO(3)$ 
\cite{hamermesh64} spanned by
$\hat{B}_{1,M}$, $M=0,\pm 1$.
The Casimir-operator of  $SU(2l+1)$ is
\begin{equation}
\hat{C}_u=\sum_{L=1}^{2l}(-1)^L \hat{B}_L^2
\end{equation}
and that of $SO(2l+1)$
\begin{equation}
\hat{C}_o=\sum_{L=1 \atop L: \text{odd}}^{2l-1} \hat{B}_L^2.
\label{eq:codef}
\end{equation}
The hamiltonian $\hat{h}$ defined in Eq. (\ref{eq:hath}) commutes with
$\hat{L}^2$, $\hat{L}_z$, $\hat{S}^2$, $\hat{S}_z$ and with the Casimir
operator $\hat{C}_u$. In fact, the Wigner-$3j$ symbol vanishes for 
$L$ odd, thus in Eq. (\ref{eq:hath}) the sum over $L$ runs over even
values of $L$. In the special case $l=1$, because the operator
$\hat{B}_2^2=\hat{C}_u-\hat{C}_o$, and in case $l=2$, because of the
accidental coincidence
\[
\left(\begin{array}{ccc}
2 & 2 & 2 \\
0 & 0 & 0
\end{array}\right)^2=
\left(\begin{array}{ccc}
2 & 2 & 4 \\
0 & 0 & 0
\end{array}\right)^2=\frac{2}{35},
\]
$\hat{h}$ can be expressed entirely in terms of $N$, $\hat{C}_u$ and 
$\hat{C}_o$. For $l>2$ this is not true anymore, and furthermore
$\hat{C}_o$ does not commute with $\hat{h}$.

However, let us consider the model-Hamiltonian
\begin{equation}
\hat{h}_m=\frac{[l]}{8}\hat{N}-\frac{\hat{N}^2}{8}-\frac{[l]}{4(2l+3)}
\left(\hat{C}_u- \hat{C}_o \right),
\label{eq:h_mdef}
\end{equation}
This model hamiltonian has the following important properties:
In the special cases $l=1,2$ the two Hamiltonians $\hat{h}$ and $\hat{h}_m$
agree.
For $l>2$ the two hamiltonians are different and we write for the 
difference
\begin{equation}
\hat{R}\equiv \hat{h}-\hat{h}_m=
\sum_{L=2 \atop L:even}^{2l}\left(\left(\begin{array}{ccc}
l&l&L\\
0&0&0
\end{array}\right)^2-\frac{2}{(2l+3)[l]}\right)
\hat{B}_L^2.
\end{equation}
In the next sections we show that the averaged value of $\hat{R}$ with the
ground states of $\hat{h}_m$ is zero
if the
particle number equal to 
\begin{equation}
N_n=n(2s+1), \quad n=0,1,\ldots,(2l+1).
\end{equation}
i.e., N is a multiple of $(2s+1)$. In \cite{Csordas04} we have reported that
a clusterization in the wave function occur at particle numbers $N_n$ 
and that the ground state energy per particle has local minima at the 
$n$-cluster particle number $N_n$. We also show that $\hat{h}$  and $\hat{h}_m$
share the same ground states for even particle numbers if $s=1/2$.

\section{Cluster states}
\label{sec:clust_states}

Making a full numerical diagonalization of the hamiltonian (\ref{eq:hath})
is not easy. Fock-vectors have $(2l+1)(2s+1)$ slots, and in each
slot there is a zero or 1 due to the fermionic character of the problem.
The dimension of the full Hilbert-space is $2^{(2l+1)(2s+1)}$.
The spectra do not depend on $L_z$ and $S_z$, thus we can restrict
ourshelves to the fermionic sectors $L_z=0$, $S_z=0$ (even particle numbers)
or $L_z=0$, $S_z=1/2$ (odd particle numbers). Conserved operators
such as $\hat{S}^2$, $\hat{L}^2$, $\hat{N}$ and  $\hat{C}_u$ makes the
numerical problem block-diagonal, but still the computer time and storage
required grows exponentially fast as soon as we increase the open shell
quantum numbers $s$ or $l$. This motivates our analytical approach besides
the numerical efforts.

In \cite{Csordas04} we have investigated the ground state wave
function at the particle number $N_1$ in first quantization. In 
second quantization the corresponding wave function reads as
\begin{gather}
|1cl\rangle=\sum_k c_k S_m\left(
\prod_{i=1}^{(2s+1)}a^+_{m_i,s+1-i}
\right)|0\rangle, \nonumber \\
k\equiv (m_1,\ldots,m_{2s+1}),
\label{eq:ex1cl}
\end{gather}
with some coefficients $c_k$. Here the symbol $S_m$ is an operator
which symmetrizes its argument with respect to the indices 
$m_1,\ldots,m_{2s+1}$
(see Sec. \ref{sec:symmetrizer}).

For notational simplicity let us introduce the integer $\sigma$ by
\begin{equation}
\sigma=\frac{(2s+1)}{2}.
\end{equation}
The ground state of $h_m$, explicitly given
by
\begin{equation}
|1cl\rangle_0=S_m(\hat{Q}_{0,0}^{+\sigma})|0\rangle.
\label{eq:1clfirst}
\end{equation}
(see Sec. \ref{sec:h_mvec}).
The operator $\hat{Q}_{0,0}^{+}$ in (\ref{eq:1clfirst})
shall play a central role in the following analyzis. It
creates a pair state from the vacuum
with quantum numbers $L=S=0$ with fixed open shell quantum numbers $(l,s)$:
\begin{equation}
\hat{Q}_{0,0}=\frac{1}{\sqrt{[l][s]}}\sum_{m,\nu}(-1)^{s-\nu+l-m}
a_{m,\nu}^+ a_{-m,-\nu}^+.
\label{eq:Q00def}
\end{equation}

Special cases of (\ref{eq:ex1cl}) are
\begin{equation}
|1cl\rangle=|1cl\rangle_0, \quad
\begin{cases}
 l=1,2,\quad  s\textrm{:arbitrary } \\
s=1/2,\quad l\textrm{:arbitrary}
\end{cases}
\label{eq:exactcases}
\end{equation}
since $\hat{h}$ and $\hat{h}_m$ coincides for $l=1,2$. The equality
for $s=1/2$ will be considered in Appendix \ref{app:spinonehalf}.

The first example when (\ref{eq:exactcases}) is not valid occur for 
$l=3$, $s=3/2$. $N=N_1=4$ is the 
first-cluster particle number. The ground state is in the $L=L_z=S=S_z=0$
fermionic sector. In this sector there are five orthonormal basis vectors.
The matrix elements of $\hat{h}$ on this basis are the following
\[
h_{i,j}=\left(
\begin{array}{ccccc}
0 &0 &0 &0 &0 \\ 
0 &-\frac{14}{11} &0 &0 &0 \\ 
0 &0 &-\frac{7}{2} &0 & 0\\ 
0 &0 &0 &-\frac{4886}{2145} &-\frac{14}{3}\sqrt\frac{2}{715} \\ 
0 &0 &0 &-\frac{14}{3}\sqrt\frac{2}{715} & -\frac{35}{6}\\ 
\end{array}\right)
\]
The true ground state energy for $N=4$ comes from the lowest $2\times2$
block-diagonal and has the value
\[
E=-\frac{7(1657+\sqrt{537729})}{2860}\approx -5.85038,
\]
which is quite close to $h_{5,5}=-35/6\approx -5.83333$.
Numerically $h_{i,j}$ in the same $2\times 2$ block is
\[
h_{i,j}\approx\left( 
\begin{array}{cc}
-2.27786 & -0.246813 \\
-0.246813 & -5.83333
\end{array}
\right),
\]
and the lowest energy state belongs to the eigenvector 
\begin{equation}
v_g\approx \left(
\begin{array}{c}
0.0690865 \\ 1
\end{array}\right).
\label{eq:v_g}
\end{equation}
It is interesting to present the matrix elements of $\hat{C}_u$
\[
C(SU(7))_{i,j}=\left( 
\begin{array}{cc}
\frac{264}{7} & 0 \\
0 & \frac{264}{7}
\end{array}
\right)
\]
and that of  $\hat{C}_o$
\[
C(SO(7))_{i,j}=\left( 
\begin{array}{cc}
18 & 0 \\
0 & 0 
\end{array}
\right)
\]
There is a small admixture of two
vectors belonging to different eigenvalues of $\hat{C}_o$ and the
dominant vector-part of the eigenvector belong to such a vector,
which is an eigenvector of $\hat{C}_o$ with eigenvalue zero. 
Analyzing further this fifth vector it turns out that it is still
given by (\ref{eq:1clfirst}) with $l=3,s=3/2$. 
The average value of $\hat{h}$ by this vector is equal to that of 
$\hat{h}_m$. Similar property will be proven in Sec. \ref{sec:aver_ener}
also for multi-cluster states.
It is important to stress that the fourth vector 
and the true ground state $v_g$ (\ref{eq:v_g}) are also a symmetrized
states with $S=0$.
\begin{table}
\begin{ruledtabular}
\begin{tabular}{c|cccc}
    &s=1/2&s=3/2&s=5/2& s=7/2 \\
\hline
l=1 & -0.75 & -3.30 & -7.65  & -13.80 \\
l=2 & -1.25 & -4.64 & -10.18 & -17.86 \\
l=3 & -1.75 & -5.85 & -12.32 & -21.17 \\
l=4 & -2.25 & -6.99 & -14.26 & -24.12 \\
l=5 & -2.75 & -8.09 & -16.09 & --- \\
l=6 & -3.25 & -9.17 & -17.83 & --- \\
\hline\hline
l=1 & -0.75 & -3.30 & -7.65  & -13.80 \\
l=2 & -1.25 & -4.64 & -10.18 & -17.86 \\
l=3 & -1.75 & -5.83 & -12.25 & -21.00 \\
l=4 & -2.25 & -6.95 & -14.11 & -23.72 \\
l=5 & -2.75 & -8.04 & -15.86 & --- \\
l=6 & -3.25 & -9.10 & -17.55 & --- \\
\end{tabular}
\end{ruledtabular}
\caption{Ground state energies of $\hat{h}$
(upper part) and $\hat{h}_m$ (lower part) up to two decimal 
digits precision at the one-cluster particle numbers $N_1=(2s+1)$.
\label{table:ground}}
\end{table}

In Table \ref{table:ground}.\ we show the numerically calculated
ground state energies of $\hat{h}$ (\ref{eq:hath}) and $\hat{h}_m$ 
(\ref{eq:h_mdef}). 
From the data it is clearly seen
that for $l=1,2$ or for $s=1/2$ Eq. (\ref{eq:exactcases}) exact 
and for all the other cases $|1cl\rangle\approx |1cl\rangle_0$ is a
rather good approximation. This shows the importance of the model
hamiltonian $\hat{h}_m$. The dominating contribution to the energy
(lower part of Table \ref{table:ground}) can be written in the form
\[
\epsilon_1=-\sigma\epsilon^{(2)}-\frac{\sigma(\sigma-1)}{2}\delta,
\] 
where $\epsilon^{(2)}$ and $\delta$ are independent of the spin. Their
explicit expressions will be derived in Sec. \ref{sec:h_mvec}.
The first term  gives
the energy of $\sigma$ independent pairs while the second
term lowers this energy, indicating clearly that the cluster wave
function gives lower energy than the wave function of independent pairs.
Note that an eigenfunction of $\hat{h}_m$ exists leading to the eigenvalue
$-\sigma\epsilon^{(2)}$ if the inequality $\sigma\le 2l+1$ is fulfilled.
For spin one half particles one has only of course, the first term, the energy
of a single pair.

Both $|1cl\rangle$ and $|1cl\rangle_0$ involve the operator $S_m$,
therefore, let us study the properties of 
the symmetrizer $S_m$.

\section{The properties of the  symmetrizer $S_m$}
\label{sec:symmetrizer}

Let us introduce the symmetrizer symbol $S_m$ by the following properties:
\textit{i)} $S_m$ is linear for its argument, 
\textit{ii)}   
$S_m$ symmetrizes any pure 
operator-product $a^+_{m_1,\nu_1}\ldots a^+_{m_p,\nu_p} $
of creation operators $a^+_{m_i,\nu_i}$, $i=1,\ldots,p$ 
with respect to {\em all} $m_i$ without changing the order 
of spin projections $(\nu_1,\ldots,\nu_p)$ including 
the combinatorical normalization. As an example:
\begin{gather*}
S_m(a^+_{0,3/2}a^+_{1,1/2}a^+_{0,-1/2})=\frac{1}{3}\bigl(
a^+_{0,3/2}a^+_{1,1/2}a^+_{0,-1/2}\bigr. \\
\bigl.
+a^+_{1,3/2}a^+_{0,1/2}a^+_{0,-1/2}+a^+_{0,3/2}a^+_{0,1/2}a^+_{1,-1/2}
\bigr).
\end{gather*}
({\em After} symmetrization, of course, one can change the
order of creation operators as we wish using the anti-commutativity
of creation operators).  

Very important special case is $p=2s+1$. Writing down all the terms in
$S_m(a^+_{m_1,\nu_1}\ldots a^+_{m_{2s+1},\nu_{2s+1}})$
it turns out that it must be antisymmetric in all spin index
$(\nu_1,\ldots,\nu_{2s+1})$. Therefore, by inspection we have
\begin{gather}
S_m(a^+_{m_1,\nu_1}\ldots a^+_{m_{2s+1},\nu_{2s+1}})=
\epsilon_{\nu_1,\ldots,\nu_{2s+1}} \nonumber \\
\times S_m(a^+_{m_1,s}\ldots a^+_{m_{2s+1},-s}),
\label{eq:id1}
\end{gather}
where $\epsilon_{\nu_1,\ldots,\nu_{2s+1}}$ is the antisymmetric 
tensor with the convention $\epsilon_{s,s-1,\ldots,-s}=1$. Further
expansion is possible for the combination
\begin{gather}
S_m(a^+_{m_1,s}\ldots a^+_{m_{2s+1},-s})=\frac{1}{(2s+1)!}
\sum_{\nu_1}\ldots\sum_{\nu_{2s+1}}\nonumber\\
\times\epsilon_{\nu_1,\ldots,\nu_{2s+1}}
a^+_{m_1,\nu_1}\ldots a^+_{m_{2s+1},\nu_{2s+1}}
\end{gather}
Next, we enumerate some properties of $S_m$ useful in the following.
If
\begin{eqnarray}
\hat{B}&=&\sum_{m,n,\nu}f_{m,n}a^+_{m,\nu}a_{n,\nu} \nonumber\\
\hat{A}_\alpha&=&\sum_{m,n,\nu}(-1)^{s-\nu} g^{(\alpha)}_{m,n}
a^+_{m,\nu}a^+_{n,-\nu},\quad \alpha=1,\ldots,\sigma,\nonumber\\
g^{(\alpha)}_{m,n}&=&g^{(\alpha)}_{n,m}
\label{eq:cond1}
\end{eqnarray} 
where $f_{m,n}$ and $g^{(\alpha)}_{m,n}$ are some numbers,
then 
\begin{gather}
S_m(\hat{A}_1 \ldots \hat{A}_\sigma)=
\frac{2^\sigma \sigma!(-1)^{\frac{\sigma(\sigma-1)}{2}}}{(2\sigma)!}
\sum_{m_1,\ldots,m_{2\sigma} \atop \nu_1,\ldots,\nu_{2\sigma}}
\epsilon_{\nu_1,\ldots,\nu_{2\sigma}}
\nonumber\\
\times
g^{(1)}_{m_1,m_2}\ldots g^{(\sigma)}_{m_{2\sigma-1},m_{2\sigma}}
a^+_{m_1,\nu_1}a^+_{m_2,\nu_2}\ldots 
a^+_{m_{2\sigma},\nu_{2\sigma}}
\label{eq:id2}
\end{gather}
This identity can be proven using Eqs. (\ref{eq:id1})-(\ref{eq:id2}) and 
\begin{gather}
\sum_{\nu_1}\ldots\sum_{\nu_{\sigma}}(-1)^{s-\nu_1}\ldots(-1)^{s-\nu_\sigma}
\epsilon_{\nu_1,-\nu_1,\ldots,\nu_{\sigma},-\nu_{\sigma}}\nonumber\\
=2^\sigma \sigma!(-1)^{\frac{\sigma(\sigma-1)}{2}}.
\end{gather}
If conditions (\ref{eq:cond1}) hold then again from 
Eqs. (\ref{eq:id1})-(\ref{eq:id2}) one can derive the 
identity
\begin{equation}
\left[\hat{B},S_m(\hat{A}_1 \ldots \hat{A}_\sigma) \right]=
S_m([\hat{B},\hat{A}_1 \ldots \hat{A}_\sigma]),
\label{eq:id3}
\end{equation}
where $[\ldots]$ denote a commutator.

\section{Cluster states and ground state energies of $\hat{h}_m$}
\label{sec:h_mvec}

Let us consider the states
\begin{equation}
|ncl\rangle_0=\hat{Q}^{+n}|0\rangle, \quad n=0,\ldots,(2l+1),
\label{eq:ncl}
\end{equation}
where the operator $\hat{Q}$ is given by
\begin{equation}
\hat{Q}^+=S_m(\hat{Q}^{+\sigma}_{0,0}).
\label{eq:qdef}
\end{equation}
$\hat{Q}^+$ creates a $(2s+1)$ particle state. In the following
we shall prove that $|ncl\rangle_0$ is an eigenstate of the 
model hamiltonian $\hat{h}_m$ (\ref{eq:h_mdef}).
It is easy to show that the state $|ncl\rangle_0$
is annihilitated by the Casimir operator $\hat{C}_o$ (See Eq. (\ref{eq:codef}))
of
$SO(2l+1)$
\begin{equation}
\hat{C}_o |ncl\rangle_0 =0
\label{eq:Co_zero}
\end{equation}
by moving the operators $\hat{B}_{L,M}$ towards to the vacuum
using the identity
\begin{equation}
\left[\hat{B}_{L,M}, S_m(\hat{Q}^{+\sigma}_{0,0})\right]=
\frac{2\sigma (-1)^{L-M}}{\sqrt{[l]}}S_m(\hat{Q}^{+}_{L,M}
\hat{Q}^{+\sigma-1}_{0,0}),
\label{eq:id4}
\end{equation}
where the operators 
\begin{eqnarray}
\hat{Q}_{L,M}^+&=&\sqrt{\frac{[L]}{[s]}}\sum_{m=-l}^l\sum_{\nu=-s}^s 
(-1)^{s-\nu+M}
\nonumber\\
&& \times
\left(\begin{array}{ccc}
l&l&L\\
m&M-m&-M
\end{array}\right)
a_{m,\nu}^+ a_{M-m,-\nu}^+
\label{eq:qlmdef}
\end{eqnarray}
create also pair states with $S=0$, but with angular momentum quantum
numbers $L$ and $M$. For $L=M=0$ this expression agrees with
(\ref{eq:Q00def}).
It is easy to prove that
\[
\hat{Q}_{2p+1,M}=0, \quad p:\text{integer}.
\]
$\hat{C}_o$ is build up from $\hat{B}_{L,M}$ with $L$ odd
but the right hand side of (\ref{eq:id4}) in that case zero, because
$\hat{Q}^{+}_{L,M}=0$ for $L$ odd. As a result $\hat{C}_o$ can be moved
towards to the vacuum, which is annihilated by $\hat{C}_o$.
As a side result we have the property
\begin{equation}
\left[\hat{C}_o, \hat{Q}^+\right]=0,
\end{equation} 
i.e., the operator $Q^+$ commutes with $\hat{C}_o$.

In order to show that $|ncl\rangle_0$ is also an eigenvector 
of $\hat{C}_u$ it requires more elaborate calculations. In moving the
operators $\hat{B}_{L,M}$ towards the vacuum one encounters
several new objects from the commutators. For them one can use the identities
(See Appendix \ref{app:ops})
\begin{equation}
 \sum_{L=0}^{2l}\sum_{M=-L}^L (-1)^{M}S_m(\hat{Q}^{+}_{L,M}\hat{Q}^{+}_{L,-M}
\hat{Q}^{+\sigma-2}_{0,0})=[l]S_m(\hat{Q}^{+\sigma}_{0,0}),
\label{EQ:IMP_ID_A}
\end{equation}
and 
\begin{gather}
\sum_{L=0}^{2l}\sum_{M=-L}^L (-1)^{M}S_m(\hat{Q}^{+}_{L,M}
\hat{Q}^{+\sigma-1}_{0,0})S_m(\hat{Q}^{+}_{L,-M}
\hat{Q}^{+\sigma-1}_{0,0}) \nonumber\\
=-\frac{[l]}{[s]} S_m(\hat{Q}^{+\sigma}_{0,0})^2.
\label{EQ:IMP_ID}
\end{gather}
As a result one has
\begin{equation}
\hat{C}_u|ncl\rangle_0=\frac{n([l]-n)}{[l]}[s]([l]+[s])|ncl\rangle_0,
\end{equation}
i.e., $|ncl\rangle_0$ as given by (\ref{eq:ncl}) 
is really an eigenvector of $\hat{C}_u$. 
Correspondingly for $\hat{h}_m$ (\ref{eq:h_mdef}) using (\ref{eq:Co_zero})
one obtains:
\begin{equation}
\hat{h}_m|ncl\rangle_0=\epsilon_n|ncl\rangle_0,
\label{eq:hm_eigenvalueeq}
\end{equation}
with
\begin{equation}
\epsilon_n=-\frac{n(2l+1)(2s+1)}{8(2l+3)}
\left[ n(2s-1)+2l+1+4s\right].
\label{eq:En}
\end{equation}
This expression can be cast into the form
\begin{equation}
\epsilon_n=n\epsilon_1+\frac{n(n-1)}{2}\gamma,
\label{eq:En1}
\end{equation}
where 
\begin{equation}
\epsilon_1=-\frac{(2l+1)(2s+1)}{4(2l+3)}(3s+l)
\end{equation}
and
\begin{equation}
\gamma=-\frac{(2l+1)(2s+1)}{4(2l+3)}(2s-1).
\end{equation}
The first term on the right hand side of Eq. (\ref{eq:En1})
can be interpreted as the energy of $n$ independent one-clusters
and the second term as a kind of cluster-cluster interaction
energy. Furthermore the energy $\epsilon_1$ can be rewritten as
\begin{equation}
\epsilon_1=-\sigma\epsilon^{(2)}-\frac{\sigma(\sigma-1)}{2}\delta,
\label{eq:1clenergy}
\end{equation}
where 
\begin{equation}
\epsilon^{(2)}=\frac{2l+1}{4},\quad \delta=\frac{12}{2l+3}\epsilon^{(2)}
\end{equation}
are independent of the spin. 
For spin one-half particles $\gamma=0$ and the prefactor of $\delta$ in
(\ref{eq:1clenergy}) is zero, which means that the clusters
consist of pairs which are independent.

One can raise the question what is the ratio of the interaction energies
of two pairs within the same cluster and when they belong to two
different clusters. This ratio is equal to
\begin{equation}
\frac{\delta}{\gamma/\sigma^2}=\frac{3(2s+1)}{2s-1}, \quad s>1.
\end{equation}
It is remarkable that this expression is $l$-independent.
This ratio is always bigger than one, monotonically decreasing with
increasing $s$.

\section{Average energies in cluster states}
\label{sec:aver_ener}

Acting with the operators $\hat{B}_L^2$ (See Eqs. 
(\ref{eq:hath}),(\ref{eq:blsqr})) on the cluster states of $\hat{h}_m$
using the rules 
(\ref{eq:BQcommut}),(\ref{eq:id3})
one gets three terms:
\begin{eqnarray}
\hat{B}_L^2|ncl\rangle_0&=&\frac{4n\sigma[L]}{[l]}|ncl\rangle_0+
\frac{4n\sigma(\sigma-1)}{[l]}|\alpha_L\rangle\nonumber\\
&&+\frac{4n(n-1)\sigma^2}{[l]}|\beta_L\rangle,
\label{eq:bltoncl0}
\end{eqnarray}
where the vectors $|\alpha_L\rangle$ and $|\beta_L\rangle$
are defined as follows:
\begin{equation}
|\alpha_L\rangle=Q^{+(n-1)}\sum_{M=-L}^L (-1)^{M}
S_m(\hat{Q}^{+}_{L,M}\hat{Q}^{+}_{L,-M} \hat{Q}^{+\sigma-2}_{0,0})|0\rangle
\label{eq:alphadef}
\end{equation}
and
\begin{gather}
|\beta_L\rangle=Q^{+(n-2)}\sum_{M=-L}^L (-1)^{M} \nonumber\\
\times S_m(\hat{Q}^{+}_{L,M}\hat{Q}^{+\sigma-1}_{0,0})
S_m(\hat{Q}^{+}_{L,-M}\hat{Q}^{+\sigma-1}_{0,0})|0\rangle.
\label{eq:betadef}
\end{gather}
In the special case $L=0$ the two vectors agree with $|ncl\rangle_0$, and
for $L=2,4,\ldots,2l$ 
it can be shown that $|\alpha_L\rangle$  and $|\beta_L\rangle$ can be 
decomposed as
\begin{equation}
|\alpha_L\rangle=\frac{4[L]}{2l+3}|ncl\rangle_0+|\alpha_L^\perp\rangle,
\quad L=2,4,\ldots,2l
\label{eq:alphadec}
\end{equation} 
and
\begin{equation}
|\beta_L\rangle=-\frac{[L]([l]+[s])}{l[s](2l+3)}|ncl\rangle_0+
|\beta_L^\perp\rangle,  \quad L=2,4,\ldots,2l
\label{eq:betadec}
\end{equation}
(See Appendix \ref{app:decomp}), where $|\alpha_L^\perp\rangle$
and $|\beta_L^\perp\rangle$ are possibly zero vectors, but if not,
they are eigenvectors of $\hat{C}_o$ with positive eigenvalues, and are
automatically orthogonal to $|ncl\rangle_0$ (which is also an eigenvector of
$\hat{C}_o$ but with zero eigenvalue).

Using the above results the energy of $\hat{h}$ averaged with $|ncl\rangle_0$ 
is
\begin{equation}
\frac{\langle ncl|\hat{h}|ncl\rangle_0}{\langle ncl|ncl\rangle_0}=
\frac{\langle ncl|\hat{h}_m|ncl\rangle_0}{\langle ncl|ncl\rangle_0}=
\epsilon_n,
\label{eq:expeq}
\end{equation}
where $\epsilon_n$ is given by (\ref{eq:En}) according to 
(\ref{eq:hm_eigenvalueeq}) .

In Appendix~\ref{app:spinonehalf} in the special case $s=1/2$
we prove the stronger statement
(\ref{eq:speconehalf}), namely that 
the ground state of $\hat{h}$ is that of $\hat{h}_m$ for even
particle numbers. This provides an alternative proof within our 
framework of Racah's result \cite{Racah52}, namely, that the ground
state of spin $1/2$ particles interacting by an attractive $\delta$-function
potential consists of independent pairs.

In order to present some numerical spectra in the most suitable form
let us introduce the energies $E'$
defined as
\begin{equation}
E'=E+\frac{Ns(2l+1)}{4}.
\label{eq:eprime}
\end{equation}
(\ref{eq:eprime}) makes such a shift, 
proportional to the particle number $N$, that $E'(N=0)=E'(N=(2l+1)(2s+1))=0$,
i.e., the energies $E'$ are zero at zero and complete filling.
At the cluster numbers $N_n$ the particle number is $N=n(2s+1)$. Making the
shift as in (\ref{eq:eprime}) the ground state energies $E'$ 
of the approximate Hamiltonian $\hat{h}_m$ at the cluster number 
$N_n$ are on the curve
\begin{equation}
E'(N)=\frac{N(2l+1)(2s-1)}{8(2l+3)(2s+1)}
\left[(2s+1)(2l+1)-N \right].
\label{eq:eprimecurve}
\end{equation}

\begin{figure}
\includegraphics[height=\linewidth,angle=270]{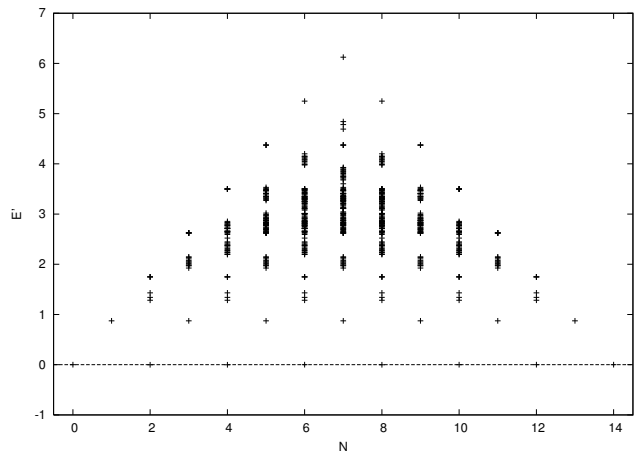}
\caption{The energy levels $E'$ of the dimensionless hamilton operator
as a function of the particle number $N$ for $l=3$, $s=1/2$. 
Individual energy levels are denoted by crosses ($+$). The dashed
line is the function (\protect\ref{eq:eprimecurve}).}
\label{fig:12_f}
\end{figure}

\begin{figure}
\includegraphics[height=\linewidth,angle=270]{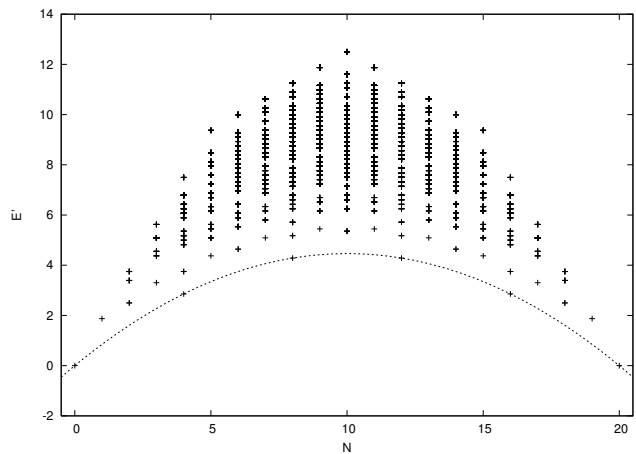}
\caption{The energy levels $E'$ of the dimensionless hamilton operator
as a function of the particle number $N$ for $l=2$, $s=3/2$. 
Individual energy levels are denoted by crosses ($+$). The dashed
line is the function (\protect\ref{eq:eprimecurve}).}
\label{fig:32_d}
\end{figure}

We show the numerically calculated ground and all excited state
energies $E'$ as a function of
the particle number $N$, for $l=3$ and $s=1/2$ in Fig.~\ref{fig:12_f}, 
and for $l=2$ and $s=3/2$ in Fig.~\ref{fig:32_d}. 

The spectra are symmetric with respect
to the point $N=(2l+1)(2s+1)/2$. This symmetry is a consequence of a
particle-hole transformation \cite{Csordas04}. 
Our data show that the 
lowest energy states at particle numbers $N_n$ are non-degenerate
and there is a gap to the first excited states at the same particle
numbers. 

The regularity in the spectrum in Fig.~\ref{fig:32_d} as compared to 
the spectrum of 
Fig.~\ref{fig:12_f} presumably arises from the fact that Fig.~\ref{fig:32_d}
refers to a situation when the hamilton operator coincides to the
model one (\ref{eq:h_mdef}) while this is not the case in Fig.~\ref{fig:12_f}.

\section{Summary and Conclusions}
\label{sec:summary}

In preceding sections we have analyzed in great details
the clusterization phenomenon 
in the one open
shell model with attractive $\delta$ interactions. We have
constructed the states $|ncl\rangle_0$ and showed that in a certain
subset of shell parameters $l,s$ these states are exact, and in other
cases they are quite good approximate ground states. We also
showed that there exist a cluster creation operator $\hat{Q}^+$
by which with repeated applications to the vacuum we can
create the (approximate or exact) ground states. 
We expressed the 
cluster creation operator $\hat{Q}^+$ in terms of the symmetrized
product of the pair creation operator $\hat{Q}_{0,0}^{+}$
by introducing the symbol $S_m$.

It is interesting to note that the symbol
$S_m$ itself, which by definition symmetrizes 
with respect to the angular momentum
indices, can be  interpreted as
an antisymmetrizer which act on spin indices.
This follows from  the form (\ref{eq:Q00otherform}) 
of the cluster creation operator
$\hat{Q}^+=S_m(\hat{Q}_{0,0}^{+\sigma})$ given by 
Eq. (\ref{eq:qdef}).

Next, let us discuss the main analytic
results (\ref{eq:expeq}). This equation
shows that in the cases $\hat{h}\ne\hat{h}_m$, i.e., when the model and
the one open shell Hamiltonian differs, $|ncl\rangle_0$ can be
regarded as a variational ansatz, i.e., the ground state energy 
of the $n$-cluster state lies
below $\epsilon_n$ given by (\ref{eq:En}). In practice $|ncl\rangle_0$
is expected to provide 
quite a good approximate ground state of $\hat{h}$
as shown in Section \ref{sec:clust_states}. for a number of
examples. Furthermore, if we make
the decomposition $\hat{h}=\hat{h}_m+\hat{R}$ and if we consider
$\hat{h}_m$ as an unperturbed Hamiltonian with unperturbed
ground state $|ncl\rangle_0$ then Eq. (\ref{eq:expeq}) shows
that we do not obtain correction to the unperturbed energy $\epsilon_n$
in first order, corrections are at least of second order,
which can be small because the off-diagonal matrix elements can
be small. This explains
the numerical finding for $l=3$, $s=3/2$ and $N=4$ that the deviation
of $\epsilon_n$ from the true ground state energy is of the order $0.3\%$
(see Sec. \ref{sec:clust_states}).
The true ground state, in general, is not the state which is 
annihilated by the Casimir operator of $SO(2l+1)$, but there
is a small admixture which mixes some other eigenstates of $\hat{C}_o$
with positive eigenvalues. Note that it remains antisymmetric in the
spin indices representing a cluster state.

The main technical problem in deriving values for the
approximate (or exact) cluster energies  $\epsilon_n$ was 
relegated to Appendix~\ref{app:decomp}. Here we applied a powerful
projection technics in a non-orthonormal and linearly non-independent
basis. This procedure showed that the unperturbed problem for
$\hat{h}_m$ is non-degenerate, if we follow how the Casimir of 
$SO(2l+1)$ acts in the relevant vector space.

\appendix
\section{Proof of the identities (\protect\ref{EQ:IMP_ID_A}) 
and (\protect\ref{EQ:IMP_ID}) }
\label{app:ops}

Let us introduce the spin dependent antisymmetric
matrix $\hat{G}_{\nu,z}$ by
\begin{equation}
\hat{G}_{\nu,z}=\sum_m (-1)^{l-m}a^+_{m,\nu}a^+_{-m,z}.
\end{equation}
If we use (\ref{eq:id2}) with 
$\hat{A}_1=\ldots=\hat{A}_\sigma=\hat{Q}_{0,0}^+$
we have 
\begin{gather}
S_m(\hat{Q}_{0,0}^{+\sigma})=
\frac{2^\sigma \sigma!(-1)^{\frac{\sigma(\sigma-1)}{2}}}
{(2\sigma)!\,\left([l][s]\right)^{\frac{\sigma}{2}}}
\sum_{\nu_1,\ldots,\nu_{2\sigma}}\epsilon_{\nu_1,\ldots,\nu_{2\sigma}}
\nonumber\\
\times \hat{G}_{\nu_1,\nu_2}\hat{G}_{\nu_3,\nu_4}\ldots 
\hat{G}_{\nu_{2\sigma-1},\nu_{2\sigma}}
\label{eq:Q00otherform}
\end{gather}
If, on the left hand side of Eq. (\ref{EQ:IMP_ID_A}) 
we use once again (\ref{eq:id2}) with
$\hat{A}_1=\hat{Q}_{L,M}^{+}$,  $\hat{A}_2=\hat{Q}_{L,-M}^{+}$
and with $\hat{A}_3=\ldots=\hat{A}_\sigma=\hat{Q}_{0,0}^+$
and perform the standard sum over $L$ and $M$ we arrive to
\begin{gather}
\sum_{L=0}^{2l}\sum_{M=-L}^L (-1)^{M}S_m(\hat{Q}^{+}_{L,M}\hat{Q}^{+}_{L,-M}
\hat{Q}^{+\sigma-2}_{0,0}) \nonumber\\
=\frac{2^\sigma \sigma!(-1)^{\frac{\sigma(\sigma-1)}{2}}[l]}
{(2\sigma)!\,\left([l][s]\right)^{\frac{\sigma}{2}}}
\sum_{m_1,n_1} (-1)^{l-m_1}(-1)^{l-n_1} \nonumber\\
\times \sum_{\nu_1,\ldots,\nu_{2\sigma}}\epsilon_{\nu_1,\ldots,\nu_{2\sigma}}
a_{m_1,\nu_1}^+ a_{n_1,\nu_2}^+ a_{-m_1,\nu_3}^+ 
a_{-n_1,\nu_4}^+\nonumber\\
\times \hat{G}_{\nu_3,\nu_4}\hat{G}_{\nu_5,\nu_6}\ldots 
\hat{G}_{\nu_{2\sigma-1},\nu_{2\sigma}}
\end{gather}
If now we change the order of $a_{n_1,\nu_2}^+$ $ a_{-m_1,\nu_3}^+$ using
anti-commutativity we can get two more $\hat{G}$ factors. In the next step
if we change the second and third indices of the $\epsilon$ tensor after
comparison with (\ref{eq:Q00otherform}) we arrive to the operator
identity Eq. (\ref{EQ:IMP_ID_A}). 

In order to proof Eq. (\ref{EQ:IMP_ID}) we proceed as above. Changing
the order of two creation operator (and dividing both sides with a common
factor) it is left to proof that
\begin{gather}
\frac{1}{[s]}\sum_{\nu_1,\ldots,\nu_{2\sigma} 
\atop z_1,\ldots,z_{2\sigma}} \epsilon_{\nu_1,\ldots,\nu_{2\sigma}}
\hat{G}_{\nu_1,\nu_2}\hat{G}_{\nu_3,\nu_4}\ldots 
\hat{G}_{\nu_{2\sigma-1},\nu_{2\sigma}}\nonumber\\
\times\epsilon_{z_1,\ldots,z_{2\sigma}}
\hat{G}_{z_1,z_2}\hat{G}_{z_3,z_4}\ldots \hat{G}_{z_{2\sigma-1},z_{2\sigma}}
\nonumber\\
=\sum_{\nu_1,\ldots,\nu_{2\sigma} 
\atop z_1,\ldots,z_{2\sigma}}\epsilon_{\nu_1,\ldots,\nu_{2\sigma}}
\hat{G}_{\nu_1,z_1}\hat{G}_{\nu_3,\nu_4}\ldots 
\hat{G}_{\nu_{2\sigma-1},\nu_{2\sigma}}\nonumber\\
\times\epsilon_{z_1,\ldots,z_{2\sigma}}
\hat{G}_{\nu_2,z_2}\hat{G}_{z_3,z_4}\ldots \hat{G}_{z_{2\sigma-1},z_{2\sigma}}.
\label{eq:tobeproven}
\end{gather}
By introducing the spin dependent matrix operator
\begin{equation}
\hat{F}_{\nu,z}=\sum_{\nu_2,\ldots,\nu_{2\sigma}}
\epsilon_{\nu,\nu_2\ldots,\nu_{2\sigma}}\hat{G}_{z,\nu_2}
\hat{G}_{\nu_3,\nu_4}\ldots\hat{G}_{\nu_{2\sigma-1},\nu_{2\sigma}}
\label{eq:Fnuz}
\end{equation}
and the spin-scalar operator
\begin{equation}
\hat{F}=\sum_{\nu_1,\ldots,\nu_{2\sigma}}\epsilon_{\nu_1,\ldots,\nu_{2\sigma}}
\hat{G}_{\nu_1,\nu_2}\hat{G}_{\nu_3,\nu_4}\ldots 
\hat{G}_{\nu_{2\sigma-1},\nu_{2\sigma}}
\end{equation}
from (\ref{eq:tobeproven}) it is left to be proven that
\begin{equation}
\frac{1}{2s+1}\hat{F}^2=\sum_{\nu,z=-s}^s\hat{F}_{\nu,z}\hat{F}_{z,\nu}
\label{eq:tobeproven1}
\end{equation}
Next, we show that
\begin{equation}
\hat{F}_{\nu,z}=\frac{\delta_{\nu,z}}{2s+1}\hat{F}.
\label{eq:FandFnuz}
\end{equation}
If this is true then Eqs. (\ref{eq:tobeproven1}), (\ref{eq:tobeproven}) 
and correspondingly (\ref{EQ:IMP_ID}) are also true. 

Let us study first the $\nu\ne z$ case in Eq. (\ref{eq:Fnuz}). In that
case among the summing indices $(\nu_2,\ldots,\nu_{2\sigma})$ of the
antisymmetric tensor $\epsilon$ the index $z$ should occur. 
Let it be the index $\nu_i=z$. This cannot
be $\nu_2$, because $\hat{G}_{z,z}=0$. It means that two {\em different}
$\hat{G}$ has the same index $z$. However, in the indices $\nu_2$ and
$\nu_{i+1}$ (if the index $z$ occur in the first index of the second 
$\hat{G}_{z,\nu_{i-1}}$) or $\nu_{i-1}$  (if the index $z$ occur in 
the second index of the second 
$\hat{G}_{\nu_{i-1},z}$ factor) the product of two $G$ factor is symmetric,
while the antisymmetric tensor $\epsilon$ in the same indices is
antisymmetric. Summing over $\nu_2$ and $\nu_{i+1}$ or $\nu_{i-1}$ we
get zero. It means, that
\begin{equation}
\hat{F}_{\nu,z}=0, \quad \textrm{if } \nu\ne z, 
\label{eq:Fnuzne}
\end{equation}
Next, let us study $\hat{F}_{\nu,\nu}$ with $\nu$ fixed. This
particular index $\nu$ should occur among the summing indices
of $\hat{F}$. It can be: $\nu_1=\nu$ or $\nu_2=\nu$, \ldots, or 
$\nu_{2\sigma}=\nu$. Changing the order of indices in all cases
such that $\nu$ be the first in the $\epsilon$ tensor and at the
same time in the first $\hat{G}$ factor 
(Using antisymmetry of the $\hat{G}$-s and $\epsilon$)
all the terms give
\begin{gather}
\hat{F}=(2s+1)\sum_{\nu_2,\ldots,\nu_{2\sigma}}
\epsilon_{\nu,\nu_2\ldots,\nu_{2\sigma}}\hat{G}_{\nu,\nu_2}
\hat{G}_{\nu_3,\nu_4}\ldots\hat{G}_{\nu_{2\sigma-1},\nu_{2\sigma}}
\nonumber\\
=(2s+1)\hat{F}_{\nu,\nu}
\label{eq:Fnuzeq}
\end{gather}
Eqs. (\ref{eq:Fnuzne}) and (\ref{eq:Fnuzeq}) together gives 
the operator identity (\ref{eq:FandFnuz}), which completes the proof
of Eq. (\ref{EQ:IMP_ID}).

\section{Decomposition of $|\alpha_L\rangle$ and $|\beta_L\rangle$}
\label{app:decomp}

Let us consider how $|\alpha_L\rangle$ or $|\beta_L\rangle$ behaves
on applying $\hat{C}_o$ to these vectors. Here we treat the calculation
for $|\beta_L\rangle$. Exactly the same method can be applied for 
$|\alpha_L\rangle$ with one minor difference, which will be shown below.

It is easy to show that
\begin{gather}
\left[\hat{B}_{L_1,M_1},\hat{Q}^+_{L_2,M_2}\right]=2\sqrt{[L_1][L_2]}
(-1)^{M_1}\sum_{L=0\atop L:even}^{2l}\sum_{M=-L}^M \nonumber\\
\times\sqrt{[L]}(-1)^M
\left(\begin{array}{ccc}
L_1&L_2&L\\
M_1&M_2&-M
\end{array}\right)
\left\{\begin{array}{ccc}
L_1&L_2&L\\
l&l&l
\end{array}\right\}Q^+_{L,M}
\label{eq:BQcommut}
\end{gather}
with $L_2$ even. 
In using Eqs. (\ref{eq:id3}) and (\ref{eq:BQcommut}) one obtains
\begin{gather}
\hat{C}_o |\beta_L\rangle=2[l](1-\delta_{L,0})|\beta_L\rangle-
4[L]\sum_{L_1=0 \atop L_1: even}^{2l} \nonumber\\
\times\left(
\frac{1}{[l]}-\left\{
\begin{array}{ccc}
L&l&l\\
L_1&l&l
\end{array}
\right\}
\right)|\beta_{L_1}\rangle
\label{eq:Coact}
\end{gather}
with $L$ even. In other words, the vector space $V$ spanned by the vectors
$\left\{|\beta_L\rangle |L=0,2,\ldots,2l \right\}$ 
is invariant under $\hat{C}_o$. If we define matrix elements in $V$ 
for an operator $\hat{O}$, for which $V$ is an invariant vector space by
\begin{equation}
\hat{O}|\beta_L\rangle= {\sum_{L'}}' O_{L',L}|\beta_{L'}\rangle,
\label{eq:Omatelem}
\end{equation}
where $\sum_{L'}'$ stands for $\sum_{L'=0, L':even}^{2l}$, the matrix elements
of $\hat{C}_o$ can be read off (\ref{eq:Coact}):
\begin{equation}
\hat{C}_o|\beta_L\rangle\equiv {\sum_{L'}}' C_{L',L}|\beta_{L'}\rangle
\end{equation}
\begin{gather}
C_{L',L}=2[l](1-\delta_{L,0})\delta_{L',L}-4[L]\left(
\frac{1}{[l]}-\left\{
\begin{array}{ccc}
L&l&l\\
L'&l&l
\end{array}
\right\}
\right)\nonumber\\
=(1-\delta_{L,0})(1-\delta_{L',0})\nonumber\\
\times\left(
2[l]\delta_{L',L}-4[L]\left(
\frac{1}{[l]}-\left\{
\begin{array}{ccc}
L&l&l\\
L'&l&l
\end{array}
\right\}
\right)
\right)
\end{gather}
for $L,L'$ even. The second equality follows  from
\begin{equation}
\left\{\begin{array}{ccc}
L&l&l\\
0&l&l
\end{array}\right\}=
\left\{\begin{array}{ccc}
0&l&l\\
L&l&l
\end{array}\right\}=\frac{1}{[l]}
\end{equation}
for $L$ even.
Actually the same matrix elements occur in vector space 
$V'=\left\{|\alpha_L\rangle |L=0,2,\ldots,2l \right\}$:
\begin{equation}
\hat{C}_o|\alpha_L\rangle\equiv {\sum_{L'}}' C_{L',L}|\alpha_{L'}\rangle.
\end{equation}

Unfortunately, neither the vectors $|\alpha_L\rangle$, nor 
the vectors $|\beta_L\rangle$ 
for $L=0,2,\ldots,2l$ are linearly independent. This clearly follows
from Eqs. (\ref{EQ:IMP_ID_A}) and  (\ref{EQ:IMP_ID}) if we apply both sides
to the vacuum and multiply from the left by an appropriate power of 
$\hat{Q}^+$. However, even in the linearly not independent (and 
correspondingly not orthonormal) basis one can still calculate matrix elements
of operator products such as
\begin{equation}
\left(\hat{O}^{(1)}\hat{O}^{(2)}\right)_{L',L}={\sum_{L''}}' O_{L',L''}^{(1)} 
O_{L'',L}^{(2)}
\end{equation}
provided $V$ is an invariant vector space of $\hat{O}^{(1)}$ and 
$\hat{O}^{(2)}$,
and furthermore, the matrix elements for $\hat{O}^{(1)}$ 
and $\hat{O}^{(2)}$ are fixed 
(This statement follows from (\ref{eq:Omatelem}) if 
$\hat{O}\equiv\hat{O}^{(2)} $ and we act on both sides with 
$\hat{O}^{(1)}$ from the left).
If one calculates matrix elements of the operator $\hat{R}$
\begin{equation}
\hat{R}=\hat{C}_o^3-(8l+6)\hat{C}_o^2+8l(2l+3)\hat{C}_o
\end{equation}
by the well-known properties \cite{Edmonds} of the Wigner $6j$-symbols 
it turns out that
\begin{equation}
\hat{R}_{L',L}=0.
\end{equation}
It also means that the operator $\hat{C}_o$ on $V$ or $V'$ fulfills
\begin{equation}
\hat{C}_o\left(\hat{C}_o-4l\hat{I}\right)
\left(\hat{C}_o-(4l+6)\hat{I}\right)=0,
\label{eq:cminpol}
\end{equation}
where $\hat{I}$ is the identity operator. 

If an operator $\hat{O}$ has the
the minimal polynomial
\begin{equation}
0=\left(\hat{O}-\lambda_1\hat{I}\right)\ldots
\left(\hat{O}-\lambda_d\hat{I}\right)
\end{equation}
with finite $d$ then $\hat{O}$ admits the decomposition
\begin{equation}
\hat{O}=\sum_{i=1}^d \lambda_i \hat{P}_i,
\end{equation}
where $\hat{P}_i$ is a projector, i.e., 
$\hat{P}_i \hat{P}_j=\hat{P}_i\delta_{i,j}$ and
\begin{equation}
\hat{P}_i=\prod_{j=1 \atop j\ne i}^d \frac{\hat{O}-\lambda_j\hat{I}}
{\lambda_i-\lambda_j}
\end{equation}

From Eq. (\ref{eq:cminpol}) is clear that on $V$ at most we have $d=3$,
and the three eigenvalues of $\hat{C}_o$ are $\lambda_1=0$, 
$\lambda_2=4l$ and $\lambda_3=(4l+6)$ respectively.
Our main purpose is to calculate the orthogonal projection of 
$|\beta_L\rangle$ to $\hat{Q}^{+n}|0\rangle=|\beta_0\rangle$. This latter
vector is an eigenvector of $\hat{C}_o$ with eigenvalue $0$, thus let 
us consider the projector of $\hat{P}_1$:
\begin{equation}
\hat{P}_1=\frac{\left(\hat{C}_o-4l\hat{I}\right)
\left(\hat{C}_o-(4l+6)\hat{I}\right)}{4l(4l+6)}.
\end{equation}
Taking matrix elements on both sides is easy. Straightforward calculation leads
to
\begin{equation}
\left(\hat{P}_1\right)_{L',L}=\delta_{L,0}\delta_{L',0}+
\frac{[L]}{l(2l+3)}(1-\delta_{L,0})(1-\delta_{L',0})
\end{equation}
If we consider the decomposition
\begin{equation}
|\beta_L\rangle=\hat{P}_1|\beta_L\rangle+(\hat{I}-\hat{P}_1)|\beta_L\rangle
\label{eq:betadecomp}
\end{equation}
the first term $\hat{P}_1|\beta_L\rangle$ is vector which is an eigenvector
of  $\hat{C}_o$ with eigenvalue zero, the second term, if it is nonzero, 
however belong to the subspace in which $\hat{C}_o$ has positive eigenvalues.
Thus, the to vectors on the right hand side of (\ref{eq:betadecomp})
are orthogonal to each other. 
Most important is the first term. Knowing the matrix elements of $\hat{P}_1$
it reads as
\begin{equation}
\hat{P}_1|\beta_L\rangle=\delta_{L,0}|\beta_0\rangle+
\frac{[L](1-\delta_{L,0})}{l(2l+3)}\sum_{L'=2 \atop L':even}^{2l}
|\beta_{L'}\rangle
\end{equation}
Equation (\ref{EQ:IMP_ID}) implies (by applying both side to the vacuum and
multiplying by $\hat{Q}^{+(n-2)}$ from the left)
\begin{equation}
\sum_{L'=2 \atop L':even}^{2l} |\beta_{L'}\rangle=
-\frac{[l]+[s]}{[s]}|\beta_{0}\rangle
\label{eq:secsum}
\end{equation}
Putting Eqs. (\ref{eq:betadecomp})-(\ref{eq:secsum}) together we obtain
the anticipated result of Eq. (\ref{eq:betadec})

To prove the corresponding result (\ref{eq:alphadec}) 
for the decomposition of $|\alpha_L\rangle$
we can proceed as above, but instead of (\ref{eq:secsum}) we should use
\begin{equation}
\sum_{L'=2 \atop L':even}^{2l} |\alpha_{L'}\rangle=
2l|\alpha_{0}\rangle,
\label{eq:firsum}
\end{equation}
which follows from the operator identity (\ref{EQ:IMP_ID_A})
if we apply both sides to the vacuum and multiply by 
$\hat{Q}^{+(n-1)}$ from the left.

\section{The special case $s=1/2$}
\label{app:spinonehalf}

In the special case $s=1/2$ the parameter $\sigma$ 
is equal to 1 and because of the only
two possibilities for $\nu=1/2=\uparrow$ or $\nu=-1/2=\downarrow$
there is no need to symmetrize $\hat{Q}^+_{L,M}$ as defined in
(\ref{eq:qlmdef})
\begin{equation}
S_m(\hat{Q}^+_{L,M})=\hat{Q}^+_{L,M}.
\end{equation}
Consequently, according to (\ref{eq:qdef})
\begin{equation}
\hat{Q}^+=\hat{Q}^+_{0,0}.
\label{eq:spec01}
\end{equation}
By definition Eq. (\ref{eq:qlmdef}) gives for $s=1/2$
\begin{eqnarray}
&&\hat{Q}^+_{L,M}=\sqrt{2[L]}(-1)^M \nonumber\\
&&\times\sum_{m=-l}^l \left(\begin{array}{ccc}
l&l&L\\
m&M-m&-M
\end{array}\right) a^+_{m,\uparrow}a^+_{M-m,\downarrow}.
\label{eq:spec02}
\end{eqnarray}
Acting with $\hat{h}$ to $|ncl\rangle_0$ Eqs. (\ref{eq:hath}), 
(\ref{eq:bltoncl0}) and (\ref{eq:En}) for $s=1/2$ leads to
\begin{equation}
\hat{h}|ncl\rangle_0=\epsilon_n|ncl\rangle_0-
\frac{[l]}{2}n(n-1)\sum_{L=0}^{2l}
\left(\begin{array}{ccc}
l&l&L\\
0&0&0
\end{array}\right)^2|\beta_L\rangle.
\end{equation}
Next we show that the second term is identically zero for $s=1/2$.
Using (\ref{eq:spec01}) (\ref{eq:spec02}) and  Eq. (\ref{eq:fmekfirst}) it
is easy to check that
\begin{gather}
\sum_{L=0}^{2l}\left(\begin{array}{ccc}
l&l&L\\
0&0&0
\end{array}\right)^2|\beta_L\rangle= \frac{8\pi}{[l]^2}\hat{Q}_{0,0}^{+(n-2)}
\sum_{m_1,m2 \atop m_3,m_4}(-1)^{m_1+m_2}
\nonumber\\
\times
f_{m_1,m_2;m_3,m_4}a^+_{m_1,\uparrow}a^+_{m_2,\downarrow}
a^+_{-m_3,\uparrow}a^+_{-m_4,\downarrow}|0\rangle
\end{gather}
If we insert into this equation the second form (\ref{eq:fmeksecond})
for $f$ we arrive at the rather long form
\begin{gather}
\sum_{L=0}^{2l}\left(\begin{array}{ccc}
l&l&L\\
0&0&0
\end{array}\right)^2|\beta_L\rangle=2\hat{Q}_{0,0}^{+(n-2)}
\sum_{m_1,m2 \atop m_3,m_4}(-1)^{m_1+m_3} \nonumber\\
\times
\sum_{L=0 \atop L:\textrm{even}}^{2l}[L]
\left(\begin{array}{ccc}
l&l&L\\
0&0&0
\end{array}\right)^2\sum_{M=-L}^L
\left(\begin{array}{ccc}
l & l & L \\
m_1 & m_3 & -M
\end{array}\right)\nonumber\\
\times\left(\begin{array}{ccc}
l & l & L \\
-m_4 & -m_2 & -M
\end{array}\right)a^+_{m_1,\uparrow}a^+_{m_2,\downarrow}
a^+_{m_3,\uparrow}a^+_{m_4,\downarrow}|0\rangle
\end{gather}
The operator $a^+_{m_1,\uparrow}a^+_{m_2,\downarrow}
a^+_{m_3,\uparrow}a^+_{m_4,\downarrow}$ is antisymmetric in 
$m_1,m_3$, while its coefficient is symmetric in the same indices,
thus, the right hand side is zero. this completes the proof
that the state $|ncl\rangle_0$ is an eigenstate of $\hat{h}$:
\begin{equation}
\hat{h}|ncl\rangle_0=
\epsilon_n|ncl\rangle_0,\quad s=1/2,
\label{eq:speconehalf}
\end{equation}
with 
\begin{equation}
\epsilon_n=-\frac{n(2l+1)}{4},\quad s=1/2.
\end{equation}

\begin{acknowledgments}
The present work has been partially supported by the Hungarian Research
National Foundation under Grant Nos. OTKA T046129 and T038202.
\end{acknowledgments}

\end{document}